\documentclass[paper,11pt]{JHEP}
\usepackage[centertags]{amsmath}
\usepackage{amsfonts} \usepackage{amssymb} \usepackage{amsthm}
\usepackage{graphicx}
\usepackage{cite}

\def\one{{\rm 1\kern -.9mm l}}                             %

\def\beq{\begin{equation}}
\def\eeq{\end{equation}}
\def\beqa{\begin{eqnarray}}
\def\eeqa{\end{eqnarray}}
\newcommand{\eqa}{\begin{eqnarray}}
\newcommand{\ena}{\end{eqnarray}}

\newcommand{\Z}{\mathbb{Z}}

\newcommand{\oo}{\mathcal{O}}
\newcommand{\D}{\Delta}

\title{Three-dimensional  quantum gravity \\according to  ST  modular bootstrap}
\author{ Ferdinando Gliozzi \\
Dipartimento di Fisica, Universit\`a di Torino\\
and Istituto Nazionale di Fisica Nucleare - sezione di Torino \\
Via P. Giuria 1, I-10125 Torino, Italy}

\abstract{
 We combine the large-$c$ ST modular bootstrap equations with the Cardy formula for the asymptotic growth of the density of states to prove that any $2d$ unitary, compact, conformal field theory (CFT) with no higher spin conserved currents leads to conflicting inequalities whenever the  entire spectrum of non-trivial  primaries lies above the BTZ threshold. As a consequence, the holographic dual of $3d$ pure gravity, if it exists, cannot be a $2d$ CFT. Consistent solutions of   ST bootstrap equations require additional primaries lying below the BTZ threshold. The lowest non-trivial primary necessarily  has  odd spin.}
\begin{document}

\section{Introduction}
\label{sec:intro}

 One  of the oldest precursors of the AdS/CFT correspondence \cite{Maldacena:1997re} is the discovery by Brown and Henneaux \cite{Brown:1986nw} that any consistent quantum theory of gravity on $AdS_3$  is a two-dimensional conformal field theory (CFT) with central charge $c=\frac3{2 G\sqrt{-\Lambda}}$ with $G$ being Newton's constant and $\Lambda$ the cosmological constant. This CFT lives on the boundary at spatial infinity and according to holographic duality encodes all the properties of the quantum theory of gravity in the bulk. In particular the $ AdS_3$ vacuum corresponds to the CFT identity module and the black hole solutions discovered by Banados Teitelboim and Zanelli (BTZ)  \cite{Banados:1992wn,Banados:1992gq}  correspond to non-trivial primaries. 

  In pure quantum gravity, owing to the absence of local degrees of freedom in the bulk, these primaries, including the identity which describes its edge modes, would form the expected complete spectrum of the corresponding  CFT.

In the following we will be interested  in a semi-classical description, which requires that the cosmological constant is small in Planck units, or equivalently that  the central charge $c$ is large \cite{Witten:2007kt}. In this limit the BTZ black holes have a horizon of positive length $A$ and a corresponding Bekenstein-Hawking entropy $ A/4G$ which nicely coincides \cite{Strominger:1997eq} with the Cardy formula  \cite{Cardy:1986ie} for the asymptotic growth of the number of states in any unitary CFT. Thus, according to holographic duality, the spectrum of heavy states of any consistent $3d$ quantum gravity is universal and corresponds to heavy black holes \cite{Hartman:2014oaa}.

BTZ black holes are locally equivalent to the  $AdS_3$ vacuum, however they are inequivalent globally since they can be obtained from the true vacuum by discrete identifications of points. As a consequence they cannot be smoothly deformed into the vacuum. We therefore expect  a gap in the mass spectrum of  pure gravity as also predicted from the requirement of  locality in the bulk theory \cite{Heemskerk:2009pn}.  Actually BTZ black holes can exist only above a mass threshold which corresponds in the large-$c$ limit to a primary of scaling dimensions $\Delta_{BTZ}=\frac{c}{12} + O(c^0)$. Prior experience with holographic duality would suggest that 
pure quantum gravity  admits a quantum completion only if one could find a large-$c$ unitary theory in which the scaling dimensions $\Delta$ of all non-trivial  primaries obey the condition 
$\Delta>\Delta_{BTZ} $. If however  such a CFT does not exist, as we will argue in this paper,  we can only conclude that the holographic dual of $AdS_3$ pure gravity, if it exists, is not a $2d$ CFT.

We consider, as usual, Euclidean $AdS_3$ whose conformal boundary geometry is a torus. The ensuing modular invariance under $PSL[2,\Z]$ and the holomorphy of the  partition function of the CFT living on the boundary are sources of important information on the properties and  consistency of the bulk theory. 

What is the meaning of modular invariance on the gravity side? In a path-integral approach to pure quantum gravity one has to sum over different $AdS_3$ geometries with fixed asymptotic boundary conditions,  i.e.  Euclidean BTZ black holes. On the CFT side this sum amounts to a regularized sum  over modular group images  (i.e. a Poincar\'e sum) of the Virasoro character of  the  identity \cite{Dijkgraaf:2000fq,Manschot:2007ha,Maloney:2007ud,Keller:2014xba,Benjamin:2020mfz}. The quantity  $Z_{MWK}$ obtained this way, known as the  Maloney, Witten and Keller  partition function,  is finite and modular invariant by construction, however suffers of some unphysical features. In particular it does not posses a discrete spectrum and some states have a negative norm, both in regime of large spin $j$ \cite{Benjamin:2019stq} and at finite $j$  \cite{Alday:2019vdr}. Different solutions have been proposed to cure this lack of unitarity, by adding some other kind of matter to the BTZ spectrum  \cite{Keller:2014xba,Benjamin:2020mfz,Benjamin:2019stq}. All these scenarios point toward the non-existence  of pure quantum gravity  as a consistent quantum theory.

A different line of thought, first advocated by Hellerman \cite{Hellerman:2009bu}, only assumes a discrete spectrum and 
modular invariance of the partition function, regarded as a smooth function of the modular parameter $\tau$  of the torus. The core of this method, known as modular bootstrap, relies on the observation that there are special values of $\tau$ where the partition function is smooth only if its derivatives fulfill specific linear constraints. The resulting infinite set of equations are known as modular bootstrap equations.  The special values are the points of the upper half-plane $H_+$ which are left invariant under the action of some non-trivial subgroup of  $PSL(2,\Z)$. Its fundamental domain accommodates three points of this kind, namely the $\Z_2$ elliptic point at $\tau=i$ , left invariant by the modular inversion $ S:\tau\to -1/\tau$; the parabolic point, or cusp, at $\tau=i\infty$, stabilized by the modular translation $T: \tau\to \tau+1$ ; and  finally the $\Z_3$ elliptic point at $\tau=\exp(2i\pi/3)$, stabilized by $ST: \tau\to -\frac1{\tau+1}$.

As first pointed out by Hellerman, the equations  associated with $\tau=i$  ($S$ bootstrap)   yield an upper bound for the allowed scaling dimension $\Delta_1$ of the first non-trivial primary of any unitary CFT with $c>1$.  Hellerman's upper bound reads $\Delta_1 < c/6+0.4737$. It has since been improved numerically as well as analytically 
\cite{Friedan:2013cba,Collier:2016cls,Afkhami-Jeddi:2019zci,Hartman:2019pcd}. To date, the strongest upper bound  obtained with $S$ modular bootstrap by extrapolating large-$c$ data, computed  with the linear programming method \cite{Rattazzi:2008pe}, is 
$\Delta_1<c/9.08$  \cite{Afkhami-Jeddi:2019zci}. This is far weaker  than the sought-after $\Delta<\Delta_{BTZ}=\frac c{12 }+O(c^0)$ , necessary to question  pure quantum gravity.

Analyticity  at the cusp $\tau=i\infty$, in the large spin limit and for $c>1$, yields  the upper bound 
$ \Delta-|j| <\frac{c-1}{12}$ on the twist gap \cite {Collier:2016cls,Benjamin:2019stq} .

$ST$ modular  bootstrap  \cite{Qualls:2014oea,Gliozzi:2019ewk,Ashrafi:2019ebi} introduces a new scale in the spectrum of primaries of any CFT. It turns out that in the bootstrap equations the scaling dimensions $\Delta_A$ of any non-trivial primary $A$ always appears in the combination $\Delta_A-\Delta_+$ with
\beq
\Delta_+= \frac{c-1}{12}+\frac1{2\sqrt{3}\pi}=\Delta_{BTZ}+O(c^0)
\label{delta}
\eeq
In the large-$c$ limit the $ST$ modular  bootstrap equations simplify dramatically. Excluding higher spin conserved currents they simply read \cite{Gliozzi:2019ewk}
\beq
\label{equations}
\sum_A (-1)^{j_A}\, N_A \,e^{-\sqrt{3}\pi\Delta_A}\left(\frac{\D_A-\D_+}{\D_+}\right) ^n=\gamma^2\,(-1)^{n+1}+O(1/\D_+)\;\;\;(n=1,2,\dots)
\eeq
where the sum is 
over the non-trivial primaries, $N_A$ is their multiplicity, $j_A$ their spin and $v= 1+e^{-\sqrt{3}\pi}$.

In this paper we will describe some important consequences of these equations. We assume their convergence and a discrete spectrum (i.e. a compact CFT) which we organize in an ascending order
$$\Delta_1<\Delta_2<\D_3<..$$ 

A salient observation is that the series in (\ref{equations})  are endowed with an intrinsic cut-off, namely $ N_A$ is bounded by the Cardy  formula for large enough $\Delta_A$ \cite{Hartman:2014oaa}. It follows that  the terms of the series with $\Delta_A>4\Delta_+=\frac c3+O(c^0)$ decay exponentially with $c$, giving rise to a further, substantial simplification of (\ref{equations}).

By exploiting these new  equations we  obtain a simple expression for the multiplicity of the light spectrum in the large-$c$ limit. 
In particular, it turns out  that $\Delta_+$,  which appears to be the CFT counterpart of the BTZ threshold, behaves as a forbidden level. To rephrase it in an apparently different way, a primary with 
$\Delta_A=\Delta_+$ decouples from the other states, as (\ref{equations}) clearly shows.

Notice that a scalar state with $\Delta=\Delta_+=\Delta_{BTZ}+O(c^0)$ corresponds on the gravity side to a massless BTZ black hole. It has a vanishing near horizon, hence a vanishing Bekenstein-Hawking entropy  \cite{Banados:1992wn,Banados:1992gq}. As a consequence its degeneracy is expected to be of order of 1. In the MWK partition function $Z_{MWK}$  it has a degeneracy of -6, one of the unphysical features of  $Z_{MWK}$  \cite{Benjamin:2020mfz}. $ST$ bootstrap avoids the issue by setting  $N_{\Delta_+}=0$.

We will obtain a general expression for the multiplicity  $N_i$ in terms of the scaling dimensions of all the primary operators $\oo_j$ with
$\D_j\le 4\D_+=c/3+O(c^0)$, namely 
\beq
 \label{main}
N_i=\gamma^2(-1)^{j_i}e^{\sqrt{3}\pi\D_i}\D_+\left(\frac{\prod_{j\not=i}\D_j}{(\D_i-\D_+)\prod_{j\not=i}
(\D_j-\D_i)}+O(1/\D_+)\right)\,,\,\D_j\le4\D_+
\eeq 
There are both positive and negative factors in the denominator of (\ref{main}) .  Requiring $N_i\ge0$ entails the following curious selection rule 
\beq
\label{rule}
\frac{1+{\rm sign}(\Delta_i-\Delta_+)}2+i+j_i={\rm even\,\, integer} .
\eeq
This gives information on the parity of the spins  in the ordered sequence of primaries. For instance, if the 
lightest primary has odd spin, then (\ref{rule}) implies $\Delta_1<\Delta_+$; if the subsequent primary has even spin then  $\Delta_2<\Delta_+$ and so on. The possible solutions of the selection rule are  encoded in the number $\kappa$ of primaries whose dimensions lie below the threshold.  The parity of spins alternates (then $i+j_i$ is even for $i\le \kappa$) always  beginning  with an odd spin. 

If $\kappa=0$ then $j_1$ is even, hence all primaries lie above the threshold $\Delta_+$. This  solution is the only one  compatible  with the
 expected spectrum of pure quantum gravity. We will show that such a solution is contradictory.  The same conclusion was already reached in \cite{Gliozzi:2019ewk}
under some restrictive conditions. Here we prove it in a more general setting.
More precisely we will demonstrate that the large-$c$ $ST$ modular  bootstrap equations, when applied to any unitary, compact CFT with no higher spin conserved currents,  generate conflicting inequalities whenever it is  assumed that the entire  spectrum of non-trivial primaries lies above the BTZ threshold. In other terms such a CFT does not exist. 

This paper is organized as follows. In the next section we combine the $ST$ bootstrap equations at large $c$ with the Cardy formula,
discuss the convergence properties of these equations, map the problem of existence of a compact CFT with a prescribed spectrum into 
a simple algebraic problem and prove our main theorem. In section 3 we obtain an expression for the multiplicity of primaries  in terms of the light spectrum of the theory. Finally in section 4 we draw some conclusions.

\section{Conflicting inequalities}
In this section we will prove that in the large-$c$ limit there are no consistent unitary compact  CFTs with no higher spin conserved currents in which the spectrum of all non-trivial primaries lies above the BTZ threshold of $c/12$.

Our starting point is the set of $ST$  modular  bootstrap equations (\ref{equations}) that we rewrite  here with some further details
\beq
\label{ST}
\sum_{i=1}^{\infty}(-1)^{j_i}N_i e^{-\sqrt{3}\pi\D_i}\left(\frac{\D_i-\D_+}{\D_+}\right)^n=v^2(-1)^{n+1}+O(1/\D_+),\,\,\,(n=1,2,\dots)\,.
\eeq
$j_i$ is the spin of the primary $\oo_i$, however only its parity matters. If $\Delta_i$ is a degenerate  level housing primaries of different spins,  we have $(-1)^{j_i}N_i=N_i^e-N_i^o$ where $N_i^e$ $(N_i^o)$ is the sum of the multiplicities of the primaries  with  even  (odd) spins. The scaling dimensions of the primary operators are arranged in an ascending order $\D_1<\D_2<\dots$.  We expand in $1/\D_+$ instead of $1/c$ because in the former case we obtain simpler expressions. We are working in the semiclassical limit $\D_+\to \infty$ and $\D_i/\D_+$ constant, thus only primaries with $\D_i\sim c$ contribute.

Let 
$S_n^{(k)}$ be the partial sum of the first $k$ terms of the $n^{\rm th}$ equation. The assumed convergence of these series means that for every $\varepsilon>0$ there exists an integer $m_n$ such that for all $k\ge m_n$ it follows that $\vert S^{(k)}_n-S_n\vert<\varepsilon$, $S_n=v^2(-1)^{n+1}+O(1/\D_+)$ being the sum of the series. If $\varepsilon$ is chosen small enough, then all the $S_n^{(k)}$'s with $k>m_n$  have the same sign as  $S_n$. In this way, we get the following \underline{\sl exact}  inequalities
\beq
\label{Sn}
S^{(k)}_{n=2\ell -1}>0\,\,\,,\,\,\,S^{(k)}_{n=2\ell }<0\,\,\,\,,\,\forall\,\,k\ge m_n\,.
\eeq
By exploiting the Cardy formula for the asymptotic growth of states we can now estimate $m_n$ and show that it is independent of $n$ for any finite $n$.

The  Cardy formula for the microcanonical entropy $S(\D)\sim\log\rho(\D)$ at large $\D$ entails the spectral density $\rho(\D)$, which for a discrete spectrum is a sum of delta functions
\beq
\rho(\D)= \sum_i N_i \,\delta(\D-\D_i)\,,
\eeq 
thus a precise definition of  $S(\D)$ requires averaging over an interval, i.e. 
$$S(\D)=\log \int_{\D-\delta}^{\D+\delta}\rho(\D')\,d\D'\,,$$
 where $\delta$ is chosen large enough to include energy levels.
In the $c\to\infty$ limit, with $\D/c$ fixed and $\D>c/6$,  it has been shown, assuming  the sparseness of the spectrum \cite{Hartman:2014oaa}, that \cite{Mukhametzhanov:2019pzy}
\beq
\label{cardy}
S(\D)=2\pi\sqrt{\frac c3\left(\D+\delta-\frac c{12}\right)}-\frac12 \log c+O(c^0)\,,
\eeq
with $\delta\sim c^\alpha$\,, $0\le\alpha<1$ and $\delta> \frac{\sqrt{3}}{\pi}$.

Actually we only need the leading behaviour of this quantity. Using the threshold $\D_+$ instead of $c$ it reads
\beq
S(\D)=4\pi \sqrt{\left(\D-\D_+\right)\D_+}\,.
\eeq
For $\D_i$ large enough, the terms of the series (\ref{ST}) behave as
\beq
(-1)^{j_i}N_i\, e^{-\sqrt{3}\pi\D_i}\left(\frac{\D_i-\D_+}{\D_+}\right)^n\sim(-1)^{j_i}e^{S(\D_i)-\sqrt{3}\pi\D_i}\left(\frac{\D_i-\D_+}{\D_+}\right)^n\,,
\eeq
therefore for any finite $n$ all the terms with $\D_i>4\,\D_+=\frac c3 +O(c^0)$ exponentially decrease with $\D_i$, so they can be all absorbed in the asymptotic symbol $O(1/\D_+)$  of (\ref{ST})  \footnote{Note that (\ref{cardy}) refers to the asymptotic multiplicity of states, while  $N_i$ refers to a proper  subset of them, that is the primaries with spins of a given parity, hence $N_i\le e^{S(\D_i)}$. Notice also that the Cardy formula has been generalized to the case of high spin values, useful in the light cone limit \cite{Kusuki:2018wpa,Collier:2018exn}. In the following we do not need this generalization. }. 

The most important consequence is that the series (\ref{ST}), for any finite $n$, are dominated by the first $m$ terms, with $\D_m\sim c/3$, thus we can replace all the $m_n$'s with $m$  in (\ref{Sn}).

The partial sum $S^{(m)}_n$  is composed of $p$ even-spin terms and $q$ odd-spin terms \footnote{In the next section we will discover that $p\sim q$ (see Fig.\ref{figure}), in contrast with the assumption $q=0$  of \cite{{Qualls:2014oea}}. Notice that 
 we use a finite number $n$ of equations for any finite $m$.} with $p+q=m$ . To simplify the notation we define
$$ S^{(m)}_n=A_n-B_n$$
with
\beqa
\label{defAB}
A_n=&&\sum_{i=1} ^p\,w_i\, a_i^n\,\,,\,B_n=\sum_{j=1} ^q\,z_j \,b_j^n\,\,\,,w_i\equiv N_i\,e^{-\sqrt{3}\pi\D_i}>0\,,\,z_j\equiv N_j\,e^{-\sqrt{3}\pi\D_j}>0\,,\cr
a_1<&&a_2<\dots<a_p\,\,,\,\,\,\,\,\,\,\,\,\,\,\,\,\,\,\,\,\,\,\,b_1<b_2<\dots<b_q\,,
\eeqa
where $a_i=\frac{\D_i-\D_+}{\D_+}$ and  $b_j=\frac{\D_j-\D_+}{\D_+}$.  We recast (a subset of) the inequalities (\ref{Sn})  in the form
\beq
\label{AB}
A_{2\ell-1}> B_{2\ell-1}\,\,,\,\,\,A_{2\ell}< B_{2\ell}\,\,\,,\,\ell=1,2,\dots
\eeq

We can now reformulate the claim at the beginning of this section, namely that there are no unitary compact CFTs without conserved higher spin currents,  in which the entire  spectrum of non-trivial primary operators lies  above the BTZ threshold, as a simple algebraic {\sl Theorem}: If the quantities $a_i$  $(i=1,2,\dots,p)$ and $b_j$  $(j=1,2,\dots,q)$ are all positive,   they cannot fulfill the first $p+q+1$ inequalities (\ref{AB}). 

The detailed  proof is fairly easy and goes as follows. We start from the following pivotal identity
\beq
\label{identity}
\sum_{j=0}^p\,\lambda_j\,A_{p-j+k}=0\,\,, (k=1,2,\dots)\,,
\eeq
which can be rewritten as
\beq
\sum_{i=1}^p\,w_i\,\left( a_i^k   \, \sum_{j=0}^p\,\lambda_j\,a_i^{p-j}\right)=0\,.
\eeq
The expression enclosed in parentheses clearly shows that the $\lambda_j$'s can be chosen to be the coefficients of the polynomial
\beq
x^k\prod_{j=1}^p(x-a_j)\equiv x^k \left(\sum_{j=0}^p\,\lambda_j\,x^{p-j}\right),
\eeq
with $\lambda_0=1$, $\lambda_1=-\sum_{i=1}^p a_i$,  $\lambda_2=\sum_{i\not=j}^p a_i\,a_j$, and so on. Since all the $a_i$'s are positive, the $\lambda_j$'s have alternating signs, therefore the inequalities (\ref{AB}), when applied to (\ref{identity}), immediately  yield 
  
  \beq
     B_{p+k}+\lambda_1\, B_{p+k-1}+\dots \lambda_p \,B_k\,\bigg\{
     \begin{matrix}
 <0\,\,{\rm if}\,p+k\,{\rm odd}\cr\,\,\, >0\,\,{\rm if}\,p+k\,
       {\rm even}\,.
\end{matrix}
     \label{Bn}
     \eeq
We take $k$ in the interval $k=1,2,\dots q+1$ and rewrite the left-hand-side (LHS) of these inequalities as
\beq
\sum_{j=1}^q\,y_j\,b_j^k\,\,\,{\rm with}\,\,\,y_j=z_j\prod_{i=1}^p\,\left(b_j-a_i\right)\,.
\eeq
Note that if all the $y_j$'s  had the same sign, then, as $k$ varies,  the sign of the  LHS would always remain  the same, contrary to  the alternating sign required by (\ref{Bn}).  We then split the $q$ variables $b_j$, in accordance with the signs of the  $y_j$'s,  into two subsets $a'_{i'}$, $(i'=1,\dots, p')$ and  $b'_{j'}$, $(j'=1,\dots, q')$ with $q=p'+q'$.  In this way we can recast (\ref{Bn}) as $q+1$ 
inequalities of the form (\ref{AB}), but with a reduced number, $q$, of variables. We repeat this process starting with new identities of the type (\ref{identity}) and further reducing  the number $q$  of variables through inequalities of type (\ref{Bn}) to $q''=q-p'$ variables  fulfilling $q''+1$ inequalities, and so on. The iteration process terminates when we are left with a single variable $b$ and the two conflicting inequalities $b>0$ and $b^2<0$. We conclude that the set of the first $p+q+1$ inequalities (\ref{AB}) does not admit any solution as long as all the variables $a_i$ and $b_j$  are positive, QED \footnote{Another intuitive proof that does not make use of the iteration process relies on the observation that by choosing the coefficients $w_i$'s small enough we could invert the sign of some inequalities (\ref{AB}), while the  inequalities (\ref{Bn}), which directly follow from (\ref{AB}), cannot change sign, as they do not depend on the $w_i$'s.}.

The question of whether this theorem can be extended to cases in which some $a_i$'s or some $b_j$'s are negative  then arises. 
The answer is simple: if some $a_i$'s are negative, that is if there are even-spin primaries lying  below the BTZ threshold,  then, provided that all the $b_j$'s are positive, the theorem holds true because deleting the negative $a_i$'s in  the inequalities (\ref{AB}) reinforces them. On the contrary if some $b_j$'s are negative the theorem is no longer true and the $ST$ bootstrap equations admit infinitely many solutions, as next section explains.

\section{Allowed spectra}

In this section we attempt to extract  useful information on the allowed spectra  of a unitary, compact CFT at large $c$ from the $ST$ bootstrap equations (\ref{equations}).

By summing each of these equations with the subsequent one, they can be recast in the form of a single inhomogeneous equation and an infinite set of homogeneous ones
\beqa
&&\sum_{i=1}^\infty w_i\,a_i-\sum_{j=1}^\infty z_j\,b_j=v^2\cr
&&\sum_{i=1}^\infty w_i\,a_i^n(1+a_i)-\sum_{j=1}^\infty z_j\,b_j^n(1+b_j)=0\,,\,(n=1,2,\dots)\,,
\eeqa
where we used the shorthand notation introduced in the previous section. These equations have the same formal structure of the bootstrap equations associated with a boundary CFT. More specifically the even-spin terms play the role of surface operators and the odd-spin 
ones the role of bulk operators in the special surface transition of the critical $3d$ Ising model. 
This system has been studied both with the extremal functional method \cite{Liendo:2012hy} as well as with the method of determinants \cite{Gliozzi:2013ysa,Gliozzi:2015qsa,Gliozzi:2016cmg}. The latter aims to obtain an approximate estimate of the low lying spectrum of the theory by looking for common zeros of the minors resulting from a suitable truncation of the above equations. In the present case it is easy to check that these minors have only trivial zeros, i.e. those giving $\D_i=\D_j$ for $i\not=j\,\,$\footnote{bootstrap equations of this kind have been described in \cite{Gliozzi:2016cmg,Gliozzi:2014jsa} under the name of non-truncatable systems. }. Similarly we do not expect that the extremal functional method used in  \cite{Liendo:2012hy} yields a solution in the present case \footnote{I thank Slava Rychkov for a useful correspondence about this point.}.

In conclusion we are unable to extract a direct information on the allowed spectrum of the theory under study. We can however find a precise relation between the low-lying spectrum of the primary operators and their multiplicity. Actually in the previous section we found that only the primaries with dimensions $\D_i\le 4\,\D_+$ can contribute to the $ST$ bootstrap equations  at  order $O(1/\D_+)$, hence we rewrite (\ref{ST}) as
\beq
\label{system}
\sum_{i=1}^p w_i\, a_i^n-\sum_{j=1}^q z_j\,b_j^n=v^2(-1)^{n+1}+O(1/\D_+)\,\,\,\,,\,p+q=m\,,\,\D_m\sim 4\D_+\,,\, n=1,2,\dots
\eeq
In (\ref{AB}) we were only interested to the sign of the partial sum of the first $m$ terms. Here we exploit a more detailed information on the value of this sum. We take into account the first $m$ equations and regard the $w_i$'s and the $z_j$'s as the unknowns. Note that the matrix of the coefficients of this linear system is a Vandermonde matrix, thus the solution can be written in a particularly simple way, namely
\beqa
\label{solab}
w_i&=& \frac{v^2}{a_i}\frac{\prod_{k\not=i}(a_k+1)\prod_{j=1}^q(b_j+1)}{\prod_{k\not=i}(a_k-a_i)\prod_{j=1}^q(b_j-a_i)}+O(1/\D_+)\cr 
z_j&=&- \frac{v^2}{b_j}\frac{\prod_{k\not=j}(b_k+1)\prod_{i=1}^p(a_i+1)}{\prod_{k\not=j}(b_k-b_j)\prod_{i=1}^p(a_i-b_j)}+O(1/\D_+)\,.
\eeqa
\FIGURE{\label{figure}
\setlength{\unitlength}{0.20mm}
\begin{picture}(350,350)
\put(50,330){a}
\put(50,310){l}\put(50,290){l}\put(50,270){o}\put(50,250){w}\put(50,230){e}\put(50,210){d}
\put(80,250){s}\put(80,230){p}\put(80,210){e}\put(80,190){c}\put(80,170){t}\put(80,150){r}\put(80,130){a}
\put(50,370){$\dots\dots\dots\dots\dots\dots\dots\dots\dots\dots\dots\dots\dots\dots$}

\put(80,330){\circle*{30}}
\put(120,330){\circle{25}}
\put(160,330){\circle*{30}}
\put(200,330){\circle{25}}
\put(240,330){\circle*{30}}
\put(280,330){\circle*{30}}
\put(320,330){\circle{25}}
\put(360,330){\circle*{30}}

\put(380,330){$\dots$}

\put(260,370){\vector(0,-1){308}}
\put(120,280){\circle*{30}}
\put(160,280){\circle{25}}
\put(200,280){\circle*{30}}
\put(240,280){\circle{25}}
\put(280,280){\circle{25}}
\put(320,280){\circle*{30}}
\put(360,280){\circle{25}}
\put(380,280){$\dots$}

\put(160,230){\circle*{30}}
\put(200,230){\circle{25}}
\put(240,230){\circle*{30}}
\put(280,230){\circle*{30}}
\put(320,230){\circle{25}}
\put(360,230){\circle*{30}}
\put(380,230){$\dots$}

\put(200,180){\circle*{30}}
\put(240,180){\circle{25}}
\put(280,180){\circle{25}}
\put(320,180){\circle*{30}}
\put(360,180){\circle{25}}
\put(380,180){$\dots$}

\put(240,130){\circle*{30}}
\put(280,130){\circle*{30}}
\put(320,130){\circle{25}}
\put(360,130){\circle*{30}}
\put(380,130){$\dots$}

\put(280,80){\circle{25}}
\put(320,80){\circle*{30}}
\put(360,80){\circle{25}}
\put(380,80){$\dots$}

\put(60,75){forbidden spectrum}
\put(230,80){\vector(1,0){50}}

\put(60,110){\line(1,0){350}}
\put(280,45){\oval(150,30)}

\put(220,39){BTZ threshold}
\end{picture}
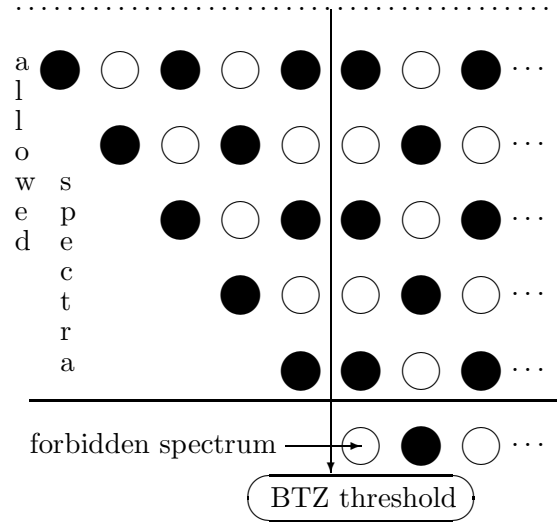
\caption{A schematic view of the allowed spectra in ST modular bootstrap at large $c$. Each row represents a solution of the selection rule (\ref{ruke}). Open (filled) circles denote even (odd) spin primaries. The bottom row is forbidden by the theorem proved in section 2.}}

Unitary CFTs require $w_i>0$ and $z_j>0$. This leads to some information on the sequence of spins and scaling dimensions of the primaries. Remember that we set the $a_i$'s and the $b_j$'s in ascending order, therefore the sign of $(a_r-a_s)$ or $(b_r-b_s)$ is the same as $r-s$, while the sign of $(a_r-b_s)$ is unknown  for the moment.  Assume for instance that the lowest primary has even spin, that is $ {\rm min}(a_1,a_2,\dots,b_1,b_2,\dots)=a_1$. Then the constraint $w_1>0$ entails $a_1\equiv\frac{\D_1-\D_+}{\D_+}>0$, hence the 
entire set of primaries lies above the BTZ threshold, a spectrum that has been excluded in the previous section.

If the lowest state is  an odd-spin primary instead, then $z_1>0$  clearly implies $\D_1<\D_+$. Moreover,  if  $\D_2<\D_+$ then 
$j_2$ is even, while if $\D_2>\D_+$,  $j_2$ is odd. Continuing in  this way it is easy to check that the general rule fixing the parity of the spin of the $n^{\rm th}$ primary is
\beq
\label{ruke}
\frac{1+{\rm sign }(\D_n-\D_+)}2+n+j_n={\rm even\,\,\,integer}\,.
\eeq
It follows that the allowed spectra are characterized by the number $\kappa$ of primaries lying below  the threshold $\D_+$. The parity of the spin alternates  both below and above $\D_+$, like in typical Regge trajectories. The lowest state always has odd spin,  and the two closest states above and below $\D_+$ have spins with the same parity (see Fig. \ref{figure}). Replacing $a_i,b_j,w_i,z_j$  in (\ref{solab}) with their definitions we get Eq.(\ref{main}) discussed in the Introduction.

The states lying below the BTZ threshold $\D_+$ cannot correspond to BTZ black holes, of course. The apparent exponential growth of their multiplicity suggests that they could be dual to black holes of a gravity theory coupled to some kind of matter.

\section{Conclusions}  

In this paper we studied some consequences of $ST$ modular  bootstrap, that is the set of equations arising from the requirement  of analyticity 
at the $\Z_3$ elliptic point of the partition function of a compact CFT formulated on a torus.  We combined these equations  in the large-$c$ limit with the Cardy formula describing the asymptotic exponential growth  of the spectral density,  in this way obtaining useful information on primaries with scaling dimensions $\D\le c/3$. In section 2 we provided a detailed proof  of a simple algebraic theorem that has important consequences for the allowed spectra of compact CFTs.  In particular, it implies  that at large $c$ any  unitary, compact CFT with no higher spin conserved currents, in which
the entire  spectrum of non-trivial  primaries lies above the BTZ threshold  leads to conflicting inequalities.  This implies
that the holographic dual of pure quantum gravity in $AdS_3$, if it exists, cannot be a unitary compact CFT. 

There is another known example of a gravity theory whose dual is not  an ordinary quantum mechanical  system on the asymptotic boundary of space-time:  Jakiw -Teitelboim (JT) dilaton gravity in two dimensions \cite{Jakiw,Teitelboim}.  It has been recently  understood  that 
the JT model is dual 
to a matrix model, a statistical ensemble of quantum mechanical systems  \cite{Saad:2019lba}. It has been very recently  suggested  that if pure $AdS_3$ gravity has a holographic dual, it could be an ensemble which generalizes matrix theory \cite{Cotler:2020ugk}.
Conversely, by averaging over a suitably defined moduli space of free $2d$ CFTs,  bulk duals have been obtained which resemble exotic theories of $3d$ gravity endowed with a large number of abelian fields  \cite{Afkhami-Jeddi:2020ezh,Mloney:2020nni,Perez:2020klz}.

Our algebraic theorem can be evaded by assuming a larger spectrum with additional primaries lying below the BTZ threshold. We found infinitely many   consistent solutions of $ST$ modular  bootstrap  equations, characterized by the number $\kappa$ of these new primaries. The lowest 
non-trivial primary necessarily has  odd spin and the spin parity of any other primary is uniquely fixed by a simple selection rule described in (\ref{rule}). We also obtained a general expression for the multiplicity of  primaries in terms of the scaling dimensions of all primaries with $\D\le c/3$ (see Eq.(\ref{main})).

In a sense,  our findings resemble those obtained with  the MWK  method  \cite{Maloney:2007ud,Keller:2014xba} used  to compute the torus partition function of pure gravity by summing over saddle points of the gravitational  path integral. Among the other unphysical features,  the resulting density of states has some negative values. This is in some way the analogue of our conflicting identities. In both cases the cure is to add new matter to the pure gravity \cite{Keller:2014xba,Benjamin:2020mfz,Benjamin:2019stq}. In our case the additional primaries lying below the BTZ threshold, in the semiclassical limit $c\to\infty$ , $\D\sim c$,  have an exponentially large multiplicity, nevertheless they cannot correspond to BTZ black holes; it would be important to understand the nature of these objects on the gravitational side.

\end{document}